\documentclass[aps,prb,twocolumn,groupedaddressfloats]{revtex4}
\usepackage{epsfig}

\begin{document}

\title[Exact]{Exact solution for the two-site 
              Kondo-Lattice Model\\
              - a limiting case for an insulator}
\author{T.~Hickel} 
\email{hickel@physik.hu-berlin.de}
\author{J.~R\"oseler}
\author{W.~Nolting}
\affiliation{
    Lehrstuhl f\"ur Festk\"orpertheorie, Institut f\"ur Physik,\\
    Humboldt-Universit\"at, 10115 Berlin}
\date{\today}

\begin{abstract}
 The Kondo-lattice model is well established as a method to describe an
 exchange coupling between single conduction electrons and localized
 magnetic moments.
 As a nontrivial exact result the zero-bandwidth limit ({\it atomic
   limit}) can be used to test approximations for this model. 
 As soon as the translational symmetry is broken (for instance by 
 sublattice structures) it is necessary to consider more than one
 lattice site. Therefore, we study as a starting point for
 generalizations the situation of a two-site cluster.
 An equation-of-motion approach is chosen to obtain the one-particle 
 Green's function. In order to determine the spectral weights of its
 energy poles, we derive different possibilities for the calculation
 of the involved correlation functions.
 In this paper the analytical exact result for the situation of
 an insulator is presented.
 In a forthcoming article we generalize the calculations to arbitrary
 electron densities, keeping as the only constraint $S=1/2$. 
\end{abstract}

\newcommand{\n}{{\hat n}^{ } }
\newcommand{\cd}[1]{c^\dagger_{#1}}
\newcommand{\cn}[1]{c^{ }_{#1}}
\newcommand{\up}{\uparrow}
\newcommand{\dn}{\downarrow}
\newcommand{\ps}{\sigma}
\newcommand{\ms}{{-\sigma}}
\renewcommand{\exp}[1]{\,{\rm exp}\!\left[ #1 \right]}
\newcommand{\ie}{{\rm i}}
\newcommand{\vz}{\epsilon}
\newcommand{\T}{\tilde T}
\newcommand{\CF}[1]{\left\langle#1\right\rangle}
\newcommand{\GFt}[2]{\left\langle\!\left\langle#1 ;#2
                     \right\rangle\!\right\rangle}
\newcommand{\GF}[2]{\left\langle\!\left\langle#1 ;#2
                    \right\rangle\!\right\rangle_E}
\newcommand{\Gf}[1]{\left\langle\!\left\langle#1
                    \right\rangle\!\right\rangle^{(\vz)}}
\newcommand{\Gff}[1]{\left\langle\!\left\langle#1
                    \right\rangle\!\right\rangle}

\newcommand{\ts}{s}
\newcommand{\os}{{\bar s}}
\newcommand{\EM}{\hat E}
\renewcommand{\a}{a}

\newcommand{\Jh}{\mbox{$\frac{J}{2}$}}
\newcommand{\fh}{\mbox{$\frac{\hbar}{2}$}}
\newcommand{\fgh}{\mbox{$\fh\Jh\,$}}
\newcommand{\pww}{r_+}    
\newcommand{\mww}{r_-}    
\newcommand{\Ppo}{\varepsilon_1}    
\newcommand{\Pmo}{\varepsilon_2}    
\newcommand{\Ppwwp}{\varepsilon_3}  
\newcommand{\Ppwwm}{\varepsilon_5}  
\newcommand{\Pmwwp}{\varepsilon_4}  
\newcommand{\Pmwwm}{\varepsilon_6}  

\maketitle
\section{Introduction}
 The idea of the Kondo lattice model goes back to Zener \cite{Zen51a}, one of 
 the first scientists who tried to give a qualitative explanation of the physical
 behaviour of transition metals. His model is based on the existence of a local 
 magnetic moment ${\bf S}_i$ per lattice site $i$ composed by electrons of partially 
 filled electron shells.
 This assumption is well fulfilled in the 3d and 4f shells of transition metals 
 and lanthanides respectively, since these shells are screened very effectively by 
 outer filled electron shells. A strong Hund's rule coupling within the shells 
 in connection with the notion of the Wigner-Eckart theorem
 leads to a well defined quantum number $S$ given by 
 \( {\bf S}_i^2 = \hbar^2 S (S+1) \).
 Due to exchange mechanisms there exists an effective coupling between the
 localized spins, which has the Heisenberg form
   \begin{equation}
     \label{eq:Heisenberg}
     {\cal H}_{f\!f} = - \sum_{i,j} J_{ij}\, {\bf S}_i \cdot {\bf S}_j .
   \end{equation}

 The electronic behaviour of these materials is determined by another class 
 of electrons, namely the 5s and 5d/6s shells of the 3d- and 4f-systems, 
 respectively. These electrons are itinerant and their 
 uncorrelated propagation through the lattice is described by the Hamiltonian 
   \begin{equation}
     \label{eq:Hopping1}
     {\cal H}_s = \sum_{i,j} \sum_\ps T_{ij} \cd{i\ps} \cn{j\ps} .
   \end{equation}
 Using second quantization, the Fermi operator $\cn{j\ps}$ annihilates an
 electron with spin $\ps$ at site $j$ whereas $\cd{i\ps}$ creates one at 
 site $i$. The hopping integrals $T_{ij}$ are connected by Fourier 
 transformation to the single electron Bloch energies
   \begin{equation}
     \label{eq:Transfer}
     T_{ij} = \frac{1}{N} \sum_{\bf k} \epsilon({\bf k})
              {\rm e}^{{\rm i} {\bf k} ( {\bf R}_i - {\bf R}_j)} .
   \end{equation}
 
 Zener pointed out that for the transition metals it is insufficient to handle 
 the magnetic and the electronic partial system independently. Instead, he suggested
 an on-site interaction of the itinerant electron spin {\boldmath$\sigma$\unboldmath}
 with the localized magnetic moments ${\bf S}_i$ of the form
   \begin{equation}
     \label{eq:DoubleEx}
     {\cal H}_{sf} = -\frac{J}{\hbar} 
     \sum_i \mbox{\boldmath$\sigma$\unboldmath}_i \cdot {\bf S}_i .
   \end{equation}

 Due to the close relation to Kondo systems, the Hamiltonian 
 \( {\cal H}_s + {\cal H}_{sf} \) is often called 
 {\it Kondo lattice model} \cite{MSN01}. 
 Another common name is {\it s-f model} or {\it s-d model} emphasizing the 
 class of described materials. In the case of manganites the same Hamiltonian
 is discussed in the limit $J \hbar \gg T_{ij}$ and called {\it double exchange model},
 based on a special mechanism for the hopping of electrons in these insulators. 
 Furthermore, one might note that the {\it periodic Anderson model} can also be 
 mapped on a Hamiltonian of the form \( {\cal H}_s + {\cal H}_{sf} \) in
 the limiting case of small charge fluctuations. 

 Sometimes, especially if higher electron concentrations are considered,
 it is reasonable to include correlations within the
 itinerant electron partial system. The simplest possibility is 
 that of an on-site Coulomb repulsion 
   \begin{equation}
     \label{eq:Coulomb}
     {\cal H}_{ss} = \frac{U}{2} \sum_i \sum_\ps \n_{i\ps} \n_{i\ms}
               = U \sum_i \n_{i\up} \n_{i\dn} . 
   \end{equation}

 We are interested in the composed Hamiltonian
   \begin{equation}
     \label{eq:CKLM}
     {\cal H} = {\cal H}_s + {\cal H}_{ss} + {\cal H}_{sf} + {\cal H}_{f\!f},
   \end{equation}
 which we shall call {\bf correlated Kondo lattice model} (CKLM). 
 By making certain assumptions for the parameters $T_{ij}, U, J_{ij}$ and $J$ 
 one might return to somewhat more special situations.

 The CKLM is not exactly soluble. One needs approximation schemes, 
 which are often difficult to justify. Limiting cases are a powerful
 tool to weaken this shortcoming. 
 
 Till now only a few exact statements on the CKLM are known.  
 One of them concerns the limiting case of a single excess electron
 in an otherwise empty conduction band, interacting with a 
 ferromagnetically saturated system of localized spins ($T = 0 {\,\rm K}$).
 Under these assumptions the electron may exist in a bound 
 state, called magnetic polaron \cite{ShM81,NMJR96}, 
 which for $J<0$ can be shown to be the ground state \cite{STU91}. 

 Another known and often used exact result for the CKLM is that of the
 zero-bandwidth limit \cite{NoM84}. 
 Here one assumes the energy dispersion to be flat:
 \( \epsilon({\bf k}) \to T_0 \), and is then able to calculate all
 possible one-particle excitation energies together with the
 corresponding spectral weights.
 Due to the relation (\ref{eq:Transfer}) the assumption is equivalent 
 to \( T_{ij} \to \delta_{ij} T_0 \), meaning that the {\it a priori}
 itinerant electrons described by \( {\cal H}_s + {\cal H}_{ss} \)
 are not allowed to change the lattice site.
 For this reason the zero-bandwidth limit describes a one-site
 situation, sometimes called {\it atomic limit}. 

 With this paper we intend to improve the atomic limit result 
 by going one step further towards the lattice. A two-site cluster 
 possesses two main advantages: First, a non-zero intersite hopping
 integral allows a motion of electrons and leads to a finite energy 
 dispersion. Compared to the flat energy band of the atomic limit, this
 is a qualitative improvement. Secondly, at least two lattice sites are
 necessary to provide a limiting case for an antiferromagnet. After
 the two-site CKLM has been solved exactly, we suppose to have a
 powerful tool to test approximation schemes for antiferromagnetic
 solutions of the full CKLM (\ref{eq:CKLM}). 
 Furthermore, we hope that the knowledge about properties and symmetries
 we gained when dealing with the two-site problem as well as the
 techniques we used to solve it are helpful tools to develop such 
 approximate schemes.  

 Especially for the Hubbard model \( {\cal H}_s + {\cal H}_{ss} \)
 there have already been some attempts to generalize exact cluster-results 
 to the lattice \cite{HaL67,Mat99,SPPL00,AMS01} and to solve even four-site
 clusters exactly \cite{Sch01}. In this context one should also mention
 efforts to perform a cluster CPA for the Hubbard model. 

 For the CKLM a complete analytical solution for the one-particle
 Green's function is so far missing. Only Matlak {\it et al.}
 \cite{MRS82} have dealt with the eigenvalue problem of a two-site cluster
 for $S=1/2$ and gave some results for certain correlation functions. 
 With our work we tried to tackle the task of the exact solution for a
 two-site cluster more comprehensively.
 The present article is devoted to the situation of an insulator, being
 characterized by an empty conduction band. It is our intention to
 explain the employed mathematical techniques and to discuss intensively
 the difficulty of a connection between cluster and lattice. 
 In a forthcoming article \cite{HRN01} we will generalize the
 calculations to arbitrary band occupations. 

\section{Cluster model}
\label{Cluster model}

\begin{figure}[ht]    
    \epsfig{file=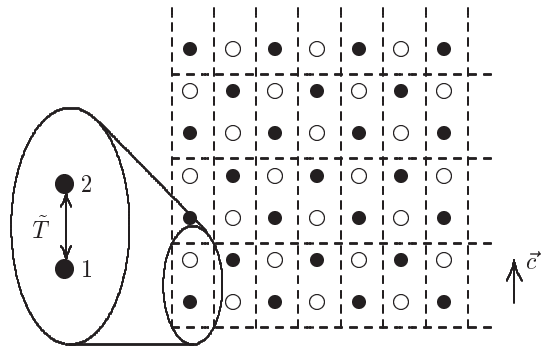, width=0.4\textwidth}
    \caption[The two-site cluster]
    {Formation of two-site clusters due to a restriction of the
     electron hopping. Different symbols for the lattice sites 
     are chosen to adumbrate an ABAB-ordered antiferromagnet.
     \label{fg:Cluster} }
 \end{figure}

 The Hamiltonian (\ref{eq:CKLM}) possesses with 
 \( {\cal H}_s \) and \( {\cal H}_{f\!f} \) two non-local terms. 
 We mentioned before that a two-site cluster has two remarkable
 properties: it is sufficient for our intension to improve the
 atomic-limit result and it is still analytically manageable.
 These desired properties guided us to treat the non-local
 terms as follows:

 The hopping integral is supposed to be 
 \[ T_{ij} = \left\{ 
               \begin{array}{ll}
                 \T \quad& i \neq j \,\mbox{and $i,j$ in same cluster,} \\
                T_0 \quad& i=j \,, \\
                  0 \quad& \mbox{$i,j$ in different clusters,}
               \end{array} 
               \right. 
 \]
 leading to a situation as shown in figure \ref{fg:Cluster}.

 In our calculations, the Heisenberg term \( {\cal H}_{f\!f} \) is  
 excluded from the Hamiltonian. This results in a quasi separation of
 the spin system and the electronic system. The calculation of the
 dynamics of localized spins is placed back in favour of a description
 of the conduction band. However, a specification of the first
 influences the latter. This offers a chance to connect the two-site
 cluster to the lattice. A localized magnetic moment, calculated within
 the cluster, would give a vanishing value. If (anti-)\-ferro\-magnetic
 substances are the subject of investigation, one should instead specify
 the values for $\CF{S^z_1}$ and $\CF{S^z_2}$ to be finite. This is
 only possible if the Hamiltonian of the cluster itself does not include a
 Heisenberg interaction of the localized spins. This assumption about
 our model is also justified since in many papers the Kondo-lattice
 model is understood to consist only of \( {\cal H}_{s}+{\cal H}_{sf} \)
 anyway \cite{MSN01}. Matlak {\it et al.} \cite{MRS82} have chosen
 another possibility by considering a mean-field approximated Heisenberg
 term as part of the Hamiltonian. Then one is confronted with even more 
 difficult mathematical expressions than given here. 

 A simplification we make in this particular article regards the number of
 electrons in the conduction band \( N \equiv \sum n_{i\ps} \). As
 mentioned before, we assume $ N=0 $ to be able to give a concise
 illustration of mathematical techniques and physical properties. 
 As a consequence, all expectation values which, like   
 $\big\langle \n_{1\up} \big\rangle$, % $\CF{\cd{2\up}\cn{1\up}}$, 
 $\big\langle S^-_2 \n_{2\up} \cd{1\up} \cn{1\dn} \big\rangle, \ldots$,
 possess in normal order an annihilation operator are zero. 
 However, a treatment with Green's functions implies the existence of 
 an additional ``test electron'', still giving rise to non-trivial results. 
 Nevertheless, the term ${\cal H}_{ss}$ of the CKLM, which describes 
 a Coulomb-interaction of {\it two} conduction electrons, is meaningless
 in the limit $N=0$. Therefore, it can be omitted. 

 Altogether, our cluster model is described by the operator
 ( \( S^\up \equiv S^+, S^\dn \equiv S^-,
      z_\up \equiv +1,  z_\dn \equiv -1 \) ):
 \begin{eqnarray}
   \bar {\cal H} &=& 
     \T \sum_{\ps=\up,\dn} 
        \left( \cd{1\ps} \cn{2\ps} + \cd{2\ps} \cn{1\ps} \right) \\
   &+& \sum_{\alpha=1}^2 \sum_{\ps=\up,\dn} \Bigg\{
                  T_0 \n_{\alpha\ps}   
              %  + \frac{U}{2} \n_{\alpha\ps} \n_{\alpha\ms}
                - \frac{J}{2} 
                  \left(z_\ps S^z_{\alpha} \n_{\alpha\ps} 
                      + S^\ps_{\alpha} \cd{\alpha\ms} \cn{\alpha\ps}  
                  \right)
               \Bigg\} . \nonumber
      \label{eq:Cluster-Ham}
 \end{eqnarray}
 Because it is a finite system, any physical quantity can in principle 
 be calculated exactly. Without any restriction for the concentration
 of conduction electrons, the dimension of the Hilbert space would be
 $4^2 \cdot (2 S +1)^2$.  We present here results for $S=1/2$. 
 For this special case one can make use of the operator identity
 \( \left( S^z_i \right)^2 \equiv \hbar^2 S^2 1\hspace{-0.55ex}{\rm l}
 \), which allows to replace any product of two spin operators at the
 same site by a single one:
 \begin{eqnarray}
     S^{\pm \sigma}_i S^{\mp \sigma}_i 
       &=& \hbar^2 S \pm \hbar z_\ps S^z_i , \\
     S^{\pm \sigma}_i S^z_i = - S^z_i S^{\pm \sigma}_i 
       &=& \mp \hbar S z_\ps S^{\pm \sigma}_i  .
 \end{eqnarray}
 Comparable relations do also exist \cite{JAG00} for $S>1/2$, giving a
 hint that the qualitative structure of our results is also valid for
 other values of $S$.

 Even so we are able to solve the eigenvalue problem for every 
 electron concentration, it is our intension to handle the cluster
 with a many-particle approach. The effort to determine the 
 complete single-particle excitation spectrum of the cluster is
 comparable for both techniques. Additionally, the calculation
 of the one-electron Green's function provides information on
 the dependence of the spectral weights not only on the model 
 parameters but also on certain expectation values as the
 averaged $z$-component of the localized spins. However, the main
 reason for our choice is the wish to get a deeper understanding
 in the physics and possible approximations of the full CKLM,
 where a many-particle treatment is inevitable.

\section{Solution for the equations of motion}
\label{Solution for the equations of motion}

 Due to the Heisenberg time dependence of the operators 
 the retarded one-particle Green's function
 \begin{equation}
   \label{eq:retGF}
   \GFt{\hat{A} (t)}{\hat{B} (t')} := - \ie \Theta( t- t') 
   \CF{\left[ \hat{A}(t) , \hat{B} (t') \right]_+ } 
 \end{equation}
 has the following equation of motion
 \begin{equation}
     \label{eq:eom}
      E \GF{\hat A}{\hat B}
         = \hbar \CF { [ \hat A, \hat B ]_+ }
         + \GF{ [ \hat A, \bar {\cal H} ]_-}{\hat B} .       
 \end{equation}
 Here, we used the Fourier-transformed notation
 \[ %\begin{equation}
   \label{eq:Fou_GF} 
   \GF{\hat A}{\hat B} := \!
   \int\limits_{-\infty}^\infty \!{\rm d}(t-t')\,
      \GFt{\hat A(t)}{\hat B(t')} \exp{\frac{\ie}{\hbar} E (t-t')}.
 \] %\end{equation}

 Motivated by the site symmetry, it is convenient to consider the 
 following combinations of Green's functions:
 \begin{eqnarray}
   \label{eq:comb1}
   \Gf{\hat C_\ts \hat D_\os \cn{\ts\ps}}
   &=& \,\,\, \GF{\hat C_\ts \hat D_\os \cn{\ts\ps}}{\cd{\ts\ps}} \nonumber\\
   &+& \vz    \GF{\hat C_\os \hat D_\ts \cn{\os\ps}}{\cd{\ts\ps}} .
 \end{eqnarray}
 Here, the site index $\os$ represents the opposite of site $\ts$
 ($\ts=1 \Rightarrow \os=2; \ts=2 \Rightarrow \os=1$), $\hat C_\ts$ and 
 $\hat D_\os$ are arbitrary products of spin operators at site $\ts$ and
 $\os$, respectively, \( \vz = + 1 \) or \( -1 \). 

 We are especially interested in the one-electron Green's function 
 \( \GF{\cn{i\ps}}{\cd{j\ps}} \) or \( \Gf{\cn{\ts\ps}} \). 
 The complete set of Green's functions generated by a repeated 
 implementation of (\ref{eq:eom}) is:\\[2ex]
 \begin{minipage}[t]{0.47\textwidth}
  \(
   \begin{array}{rclrcl}
     G_{1}^{(\vz)} &=& \Gf{\cn{\ts\ps}} \hspace{1.5cm}&
     K_{1} &=& \hbar  ,\\[0.5ex] 
     G_{2}^{(\vz)} &=& \Gf{S^z_\ts \cn{\ts\ps}} &
     K_{2} &=& \hbar \CF{S^z_\ts} ,\\[0.5ex]
     G_{3}^{(\vz)} &=& \Gf{S^z_\os \cn{\ts\ps}} &
     K_{3} &=& \hbar \CF{S^z_\os} ,\\[0.5ex]
     G_{4}^{(\vz)} &=& \Gf{S^{\ms}_\ts \cn{\ts\ms}} &
     K_{4} &=& 0 ,\\[0.5ex]
     G_{5}^{(\vz)} &=& \Gf{S^{\ms}_\os \cn{\ts\ms}} &
     K_{5} &=& 0 ,\\[0.5ex]
   \end{array}
  \)
 \end{minipage}
 \begin{minipage}[t]{0.52\textwidth}
  \(
   \begin{array}{rclrcl}
     G_{6}^{(\vz)} &=& \Gf{S^z_\ts S^z_\os \cn{\ts\ps}} &
     K_{6} &=& \hbar \CF{S^z_\ts S^z_\os} ,\\[0.5ex]
     G_{7}^{(\vz)} &=& \Gf{S^z_\os S^{\ms}_\ts \cn{\ts\ms}} &
     K_{7} &=& 0 ,\\[0.5ex]
     G_{8}^{(\vz)} &=& \Gf{S^z_\ts S^{\ms}_\os \cn{\ts\ms}} &
     K_{8} &=& 0 ,\\[0.5ex]
     G_{9}^{(\vz)} &=& \Gf{S^{\ms}_\os S^{\ps}_\ts \cn{\ts\ps}} &
     K_{9} &=& \hbar \CF{S^{\ms}_\os S^{\ps}_\ts} ,\\[0.5ex]
    G_{10}^{(\vz)} &=& \Gf{S^{\ms}_\ts S^{\ps}_\os \cn{\ts\ps}} &
    K_{10} &=& \hbar \CF{S^{\ms}_\ts S^{\ps}_\os} .
   \end{array}
  \)
 \end{minipage}
  \begin{equation}
   \label{eq:GF-list}
  \end{equation}
 In the right column the corresponding inhomogeneities in the equation of
 motion (\ref{eq:eom}) to each of the Green's functions is given. 
 Emerging Green's functions with another combination of spin operators
 or a higher number of electron operators vanish identically because of
 the constraints $S=1/2$ and $N=0$.

 It is a valuable property of the combination (\ref{eq:comb1})
 that the equations of motion of the $G^{(+)}$ and the $G^{(-)}$ Green's
 functions do not mix. Moreover, if one writes ({\ref{eq:eom}) in the
 form 
 \begin{equation}
   \label{eq:eom-matrix1}
   \sum_{j=1}^{10} m_{ij} G_j^{(\vz)} = K_i,
 \end{equation}
 one gets a matrix $M = \left(m_{ij} \right)_{i,j}$ which contains 
 $\vz$ only as a pre-factor of $\T$. If the matrix equation
 (\ref{eq:eom-matrix1}) is solved for $\vz=+1$, one obtains the
 expression for $G_i^{(-)}$ simply by replacing in $G_i^{(+)}$ the 
 parameter $\T$ by $-\T$. We therefore omit $\vz$ in the forthcoming 
 equations.

 By a couple of unitary transformations $M$ can be reduced further to blocks 
 of a size not greater than $3 \times 3$. It is favourable to introduce the
 combined Green's functions
 \begin{equation}
  \label{eq:comb2}
  \begin{array}{rcl}
   H_{0}^{(\pm)}
    &=&\mbox{$\frac{\hbar}{2}$} \Gff{\cn{\ts\ps}}
                      \pm z_\ps \Gff{S^z_\ts \cn{\ts\ps}} \\
    &&\hfill= \mbox{$\frac{1}{\hbar}$} 
              \Gff{S^{\pm\ps}_\ts S^{\mp\ps}_\ts \cn{\ts\ps}} ,\\
   H_{1}^{(\pm)}
    &=&\mbox{$\frac{\hbar}{2}$} \Gff{S^z_\os \cn{\ts\ps}}
                      \pm z_\ps \Gff{S^z_\ts S^z_\os \cn{\ts\ps}} \\
    &&\hfill= \mbox{$\frac{1}{\hbar}$}  
              \Gff{S^{\pm\ps}_\ts S^{\mp\ps}_\ts S^z_\os \cn{\ts\ps}} ,\\
   H_{2}^{(\pm)} 
    &=& \mbox{$\frac{\hbar}{2}$} H_{0}^{(\pm)} \pm z_\ps H_{1}^{(\pm)} \\
    && \hfill= \mbox{$\frac{1}{\hbar^2}$}
               \Gff{S^{\pm\ps}_\ts S^{\mp\ps}_\ts S^{\pm\ps}_\os S^{\mp\ps}_\os \cn{\ts\ps}} ,\\
   H_{3}^{(\pm)}
    &=& \mbox{$\frac{\hbar}{2}$} H_{0}^{(\mp)} \pm z_\ps H_{1}^{(\mp)} \\
    && \hfill= \mbox{$\frac{1}{\hbar^2}$}
               \Gff{S^{\mp\ps}_\ts S^{\pm\ps}_\ts S^{\pm\ps}_\os S^{\mp\ps}_\os \cn{\ts\ps}} ,\\
   H_{4}^{(\pm)}
    &=&\mbox{$\frac{\hbar}{2}$} \Gff{S^{\ms}_\ts \cn{\ts\ms}}
                      \pm z_\ps \Gff{S^z_\os S^{\ms}_\ts \cn{\ts\ms}} \\
    &&\hfill=\mbox{$\frac{1}{\hbar}$} 
             \Gff{S^{\pm\ps}_\os S^{\mp\ps}_\os S^\ms_\ts \cn{\ts\ps}} ,\\
   H_{5}^{(\pm)}
    &=&\mbox{$\frac{\hbar}{2}$} \Gff{S^{\ms}_\os \cn{\ts\ms}}
                      \pm z_\ps \Gff{S^z_\ts S^{\ms}_\os \cn{\ts\ms}} \\
    &&\hspace{9em}=\mbox{$\frac{1}{\hbar}$} 
             \Gff{S^{\pm\ps}_\ts S^{\mp\ps}_\ts S^\ms_\os \cn{\ts\ps}} .
  \end{array}
 \end{equation}
 In another step the combinations 
 \( H_3^{(-)} \pm G_{10}, H_3^{(+)} \pm G_{9} \) and
 \( H_4^{(+)} \pm H_5^{(+)} \) are used, what finishes the reduction.

 The remaining $3 \times 3$ blocks have the following matrix-structure:
 \begin{equation}
     \label{eq:Matrix1}
     M_0^{(\mu_1 \mu_2 \mu_3 \mu_4)} (\EM) = 
     \left(
       \begin{array}{ccc}
          \EM+2\mu_4 \a    & \mu_1 \T  &  0 \\
          \mu_1 \T  & \EM       & \mu_2 2\a \\
          0         & \mu_2 2\a & \EM+\mu_3 \T    
       \end{array}
     \right)
 \end{equation}
 Here, we used the abbreviation \( \a = \frac{\hbar}{2}\frac{J}{2} \)
 and \( \mu_1, \ldots, \mu_4 \) are sign parameters. The eigenvalues of 
 this matrix are
 \begin{eqnarray*}
  \EM_{01}^{(\mu_3 \mu_4)} &=& \mu_3 \T +2\mu_4 \a , \\
  \EM_{02}^{(\mu_3 \mu_4)} &=& -\sqrt{ 4\a^2 - 2\mu_3\mu_4\a\T + \T^2} , \\
  \EM_{03}^{(\mu_3 \mu_4)} &=& +\sqrt{ 4\a^2 - 2\mu_3\mu_4\a\T + \T^2} .
 \end{eqnarray*}
 For the eigenvectors one gets
 \begin{eqnarray}
     v_{01}^{(\mu_1 \mu_2 \mu_3 \mu_4)}
            &=& \frac{1}{\sqrt{3}} 
                \left(\begin{array}{c}
                  \mu_1\mu_3 \\ 1 \\ \mu_2\mu_4
                \end{array} \right) \\
     \hspace{-2em} \mbox{and for $k=2,3$} \nonumber \\
     v_{0k}^{(\mu_1 \mu_2 \mu_3 \mu_4)}
            &=& p_{k}
                \left(\begin{array}{c}
                  \mu_1 (\T-2\mu_3\mu_4\a-\mu_3\EM_{0k}) \\ 
                  -\mu_3\T+\EM_{0k} \\ 
                  2\mu_2 \a
                \end{array}\right) \\
   \mbox{with}\quad p_{k} &=& \frac{1}{2\a} 
                          \sqrt{\frac{\EM_{0k}+\mu_3\T-\mu_4\a}{3\EM_{0k}}}.
 \end{eqnarray}
 We denote by $v_{0k}[i]$ the $i$th component of $v_{0k}$ and 
 define a matrix $V$ by its components \( V_0[ik] = v_{0k}[i] \).
 Basic algebra shows that \( M_0 = V_0 \cdot D_0 \cdot V_0^{-1} \)
 with \( D_0 = {\rm diag}(\EM+\EM_{0k}) \). Using these expressions 
 one can give the components of the inverse of $M_0$ in a form 
 which is characterized by linear energy poles:
 \begin{equation}
   \label{eq:inverse}
   M_0^{-1}[ij] = \sum_{k=1}^3 v_{0k}[i] v_{0k}[j]
   \frac{1}{\EM+\EM_{0k}} . 
 \end{equation}
 This form is particularly suitable for the calculation of 
 Green's functions.

 The $10 \times 10$ matrix equation (\ref{eq:eom-matrix1}) can now be
 written in the form
 \begin{eqnarray*}
   {M_0^{(-+-+)}(E-T_0-\a)}
     \left(\!\begin{array}{c}
            H_{5}^{(-)} \\ H_{4}^{(-)} \\ H_{2}^{(-)}
     \end{array}\!\right)
   &=&\left(\!\begin{array}{c}
            R_1 \\ R_2 \\ R_3
     \end{array}\!\right) ,\\
   {M_0^{(-+-+)}(E-T_0-\a)}
     \left(\begin{array}{c}
              H_{3}^{(-)}+G_{10} \!\! \\ 
              H_{3}^{(+)}+G_{9} \\ 
              H_{4}^{(+)}+H_{5}^{(+)} \!\!
           \end{array}\right)
  &=& \left(\begin{array}{c}
              R_4+R_7 \\ R_5+R_8 \\ R_6+R_9 
            \end{array}\right) \label{eq:3Teilsystem1} ,\\
   {M_0^{(-+++)}(E-T_0-\a)}
   \left(\begin{array}{c}
            H_{3}^{(-)}-G_{10}  \!\!\\ 
            H_{3}^{(+)}-G_{9} \\ 
            H_{4}^{(+)}-H_{5}^{(+)}  \!\!
         \end{array}\right)
  &=& \left(\begin{array}{c}
            R_4-R_7 \\ R_5-R_8 \\ R_6-R_9 
          \end{array}\right) \label{eq:3Teilsystem2} 
 \end{eqnarray*}
 and solved simply by applying equation (\ref{eq:inverse}).
 The emerging inhomogeneities $R_1, \ldots, R_9$ are constructed out of
 $K_1, \ldots, K_{10}$, obeying the same rules of combination as for 
 the Green's functions. The single missing equation has an even simpler 
 structure. One can write down immediately the solution for $H_2^{(+)}$:
 \begin{eqnarray}
     H_{2}^{(+)} &=& \frac{R_0}{E-T_0-\T+\a} \\
         &=& \frac{\mbox{$\frac{\hbar}{2}$} 
                 \left( \mbox{$\frac{1}{2}$} 
                            \hbar^2 + \hbar z_\ps \CF{S^z_1} \right) 
               + \left( \mbox{$\frac{1}{2}$} 
                            \hbar^2 z_\ps \CF{S^z_2} + \hbar \CF{S^z_1 S^z_2} \right)}
                {E-T_0-\T+\a} \nonumber
 \end{eqnarray}

 To obtain an expression for the one-particle Green's function we are
 looking for, one has to take a sum of the combined functions $H_2^{(\pm)},
 \ldots H_5^{(\pm)}$. Hence,
 \begin{eqnarray}
    \label{eq:decomp}
     \Gff{\cn{\ts\ps}}
      =  G_1
     &=& \frac{1}{\hbar^2} \left( 
           H_{2}^{(+)} + H_{3}^{(+)} + H_{2}^{(-)} + H_{3}^{(-)}
         \right)  \\
     &=& \frac{1}{\hbar^2} \left( 
           H_{2}^{(+)} + H_{2}^{(-)} 
         \right) \nonumber \\
     &+&  \frac{1}{2\hbar^2}\left( \left( H_{3}^{(-)}+G_{10} \right)
                       +   \left( H_{3}^{(+)}+G_9 \right) \right) \nonumber\\
     &+& \frac{1}{2\hbar^2}\left( \left( H_{3}^{(-)}-G_{10} \right)
                       +   \left( H_{3}^{(+)}-G_9 \right) \right) . \nonumber
 \end{eqnarray}
 We now remind the fact stated after equation (\ref{eq:eom-matrix1}) that 
 this solution can be obtained equally well for $\vz = +1$ as for $\vz =-1$.
 If we add these two Green's functions 
 \begin{equation}
     \GF{\cn{\ts\ps}}{\cd{\ts\ps}} 
     = \frac{1}{2} \left(\Gff{\cn{\ts\ps}}^{(+)} + \Gff{\cn{\ts\ps}}^{(-)} \right),
 \end{equation}
 we effectively reverse the combination (\ref{eq:comb1}), which was a
 consequence of site symmetry. 

 Therefore, the following result for the one-particle Green's
 function is obtained:
 \begin{eqnarray*}
   \lefteqn{ \GF{\cn{\ts\ps}}{\cd{\ts\ps}} }\\
     &=& \left( \frac{\hbar}{ E-\Ppo } + \frac{\hbar}{ E-\Pmo } \right)
          \cdot\\&&\cdot
          \frac{1}{6 \hbar^2}  
          \left( 
             \frac{3}{2} \hbar^2
           + z_\ps \hbar \langle S^z_x \rangle
           + z_\ps \hbar \langle S^z_y \rangle 
    \right. \\&& \quad \left.
           + 2 \langle S^z_x S^z_y \rangle
           + \langle S^{-\eta}_y S^{ \eta}_x \rangle 
           + \langle S^{-\eta}_x S^{ \eta}_y \rangle 
          \right) \\
   &-& \left(  \frac{\hbar/\EM_{02}^{(-+)}}{E-\Ppwwp} 
             + \frac{\hbar/\EM_{03}^{(-+)}}{E-\Ppwwm} 
             + \frac{\hbar/\EM_{02}^{(++)}}{E-\Pmwwp} 
             + \frac{\hbar/\EM_{03}^{(++)}}{E-\Pmwwm} 
       \right) \cdot\\&&\cdot
       \frac{\hbar\Jh}{12 \hbar^2} 
       \left(  
             -2z_\ps \hbar \langle S^z_x \rangle
             + z_\ps \hbar \langle S^z_y \rangle
             +2\langle S^z_x S^z_y \rangle              
    \right. \\&& \quad \left.
             + \langle S^{-\eta}_y S^{ \eta}_x \rangle 
             + \langle S^{-\eta}_x S^{ \eta}_y \rangle
       \right)\\
   &+& \left(  \frac{\hbar}{E-\Ppwwp} 
             + \frac{\hbar}{E-\Ppwwm} 
             + \frac{\hbar}{E-\Pmwwp} 
             + \frac{\hbar}{E-\Pmwwm} 
       \right) \cdot\\&&\cdot
       \frac{1}{12 \hbar^2} 
       \left( 
               \frac{3}{2} \hbar^2
             - z_\ps \hbar \langle S^z_x \rangle
             - z_\ps \hbar \langle S^z_y \rangle
    \right. \\&& \quad \left.
             - 2 \langle S^z_x S^z_y \rangle
             - \langle S^{-\eta}_y S^{ \eta}_x \rangle 
             - \langle S^{-\eta}_x S^{ \eta}_y \rangle 
       \right) \\
   &-& \left(  \frac{\hbar/\EM_{02}^{(-+)}}{E-\Ppwwp} 
             + \frac{\hbar/\EM_{03}^{(-+)}}{E-\Ppwwm} 
             - \frac{\hbar/\EM_{02}^{(++)}}{E-\Pmwwp} 
             - \frac{\hbar/\EM_{03}^{(++)}}{E-\Pmwwm} 
       \right) \cdot\\&&\cdot
       \frac{\T}{12 \hbar^2} 
       \left(- z_\ps \hbar \langle S^z_x \rangle
             - z_\ps \hbar \langle S^z_y \rangle
             +4\langle S^z_x S^z_y \rangle              
    \right. \\&& \quad \left.
             +2\langle S^{-\eta}_y S^{ \eta}_x \rangle 
             +2\langle S^{-\eta}_x S^{ \eta}_y \rangle 
       \right) \\[-6ex]
 \end{eqnarray*}
 \begin{equation}
   \label{eq:result}
 \end{equation}
 It has the six energy poles
 \begin{eqnarray*}
     \Ppo   &=& T_0 + \a - \EM_{01}^{(-+)}
             =  T_0 - \fgh + \T ,\\ 
     \Pmo   &=& T_0 + \a - \EM_{01}^{(++)}
             =  T_0 - \fgh - \T ,\\
     \Ppwwp &=& T_0 + \a - \EM_{02}^{(-+)}
             =  T_0 + \fgh + \sqrt{ \left(\hbar\Jh\right)^2 + \hbar\Jh \T + \T^2 } ,\\
     \Pmwwp &=& T_0 + \a - \EM_{02}^{(++)}
             =  T_0 + \fgh + \sqrt{ \left(\hbar\Jh\right)^2 - \hbar\Jh \T + \T^2 } ,\\
     \Ppwwm &=& T_0 + \a - \EM_{03}^{(-+)}
             =  T_0 + \fgh - \sqrt{ \left(\hbar\Jh\right)^2 + \hbar\Jh \T + \T^2 } ,\\
     \Pmwwm &=& T_0 + \a - \EM_{03}^{(++)}
             =  T_0 + \fgh - \sqrt{ \left(\hbar\Jh\right)^2 - \hbar\Jh \T + \T^2 } ,
 \end{eqnarray*} 
 shown in figure \ref{fg:poles}.
  \begin{figure}[th]
    \begin{center}
     \epsfig{file=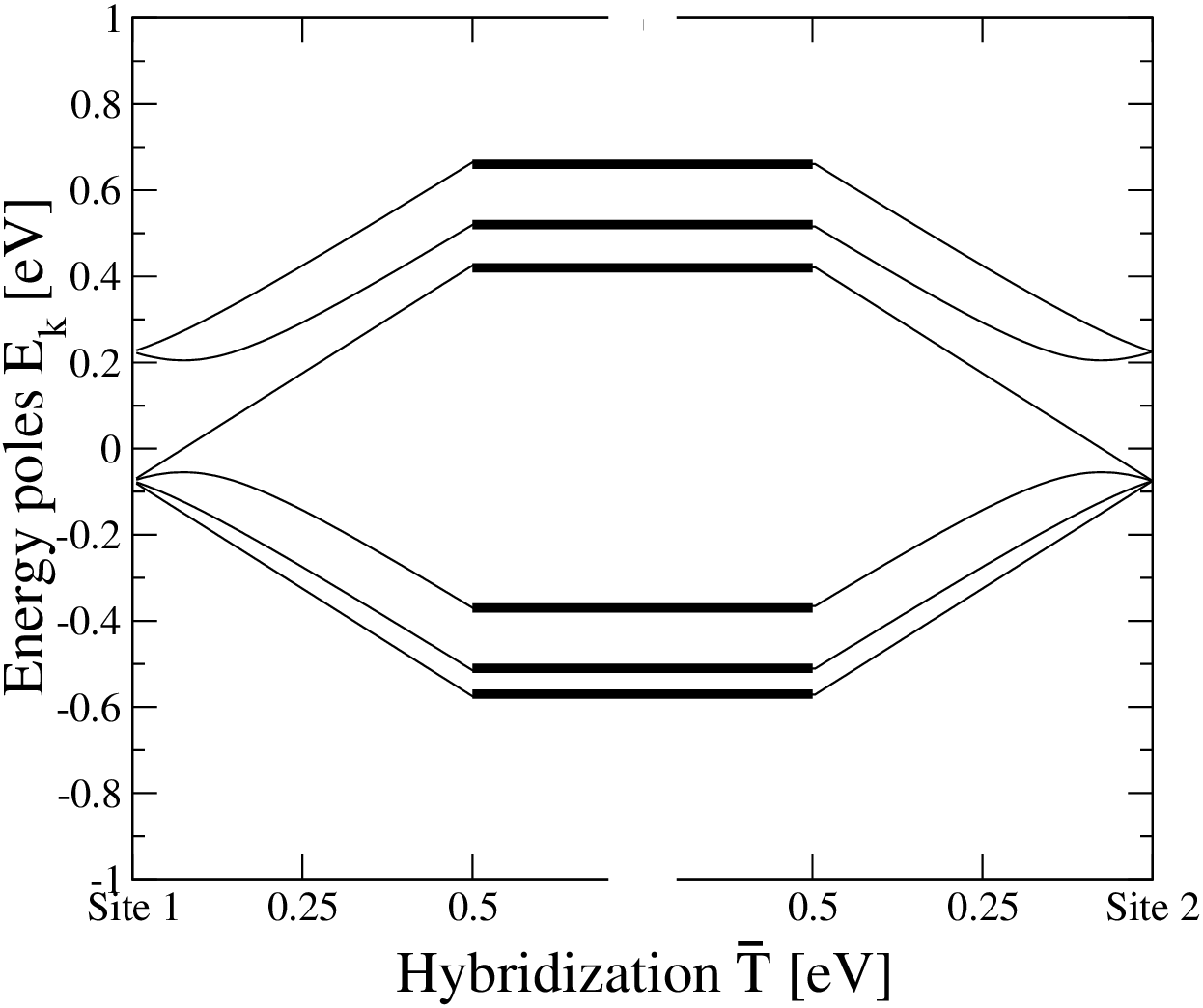, width=0.45\textwidth} 
    \end{center}
    \caption[Positions of the energy poles]
    {The hybridization $\T$ causes a splitting of the one-electron energy
     levels valid for a single site and a grouping of the excitation energies. 
     For $\T = 0.5 {\,\rm eV}$ and $J \hbar = 0.3 {\,\rm eV}, T_0 = 0 {\,\rm eV}$ 
     the arrangement of the energy poles
     \( E= \Pmo, \Ppwwm, \Pmwwm, \Ppo, \Pmwwp, \Ppwwp \) 
     (starting from the bottom) is shown.
    \label{fg:poles}}
  \end{figure}

\section{Treatment of the correlation functions}

 The result (\ref{eq:result}) provides an expression for the one-particle
 Green's function. It is characterized by six linear energy poles in the 
 denominators. They are identical with the excitation energies possible 
 if a single excess electron is placed in the otherwise empty conduction 
 band. The numerators give the spectral weights corresponding to these
 excitations. However, in contrast to the model parameters, $\T, T_0, J$, 
 which are assumed to be known, we have till now no information about
 the spin correlation functions
          \( \CF{S^z_1}, \CF{S^z_2}, \CF{S^z_1 S^z_2},   
             \CF{S^+_1 S^-_2}, \CF{S^-_1 S^+_2} \),
 leaving the spectral weights undetermined. 

 In the next two subsections we will introduce and discuss two possible concepts
 to obtain results for these correlation functions. They are especially 
 important for the treatment of a situation with $N>0$.  

 \subsection{Remaining in the cluster}

  As a result of the calculations in section \ref{Solution for the equations 
  of motion} we do not only get an expression for the Green's function 
  $G_1^{(\vz)}$ and therefore for \( \GF{\cn{\ts\ps}}{\cd{\ts\ps}} \). 
  On the contrary, we are without further effort able to write down similar
  expressions for all the other participating Green's functions
  $G_2^{(\vz)}, \ldots, G_{10}^{(\vz)}$. One just has to sum the combined 
  Green's functions $H_2^{(\pm)}, \ldots, H_5^{(\pm)}$ another way as in
  (\ref{eq:decomp}). 

  It is one possible concept to apply the well-known spectral theorem to each 
  of the determined Green's functions. If \( G_1^{(\vz)}, G_2^{(\vz)}, \ldots \) 
  or \( G_{10}^{(\vz)} \) is of the form
  \begin{equation}
    \label{eq:GF_Form}
    \GFt{\hat A}{\hat B}_{E+\ie 0^+} = \sum_{i=1}^p 
                      \frac{\hbar\,\alpha_i}{E-E_i + \ie 0^+},
  \end{equation}
  then one can derive an expression for a corresponding correlation function
  \begin{equation}
    \label{eq:spectraltheorem}
     \CF{\hat{B} \hat{A}} = \sum\limits_{i=1}^p \alpha_i \, f_-(E_i) 
          \quad = \sum\limits_{i=1}^p 
                    \frac{\alpha_i}{ {\rm e}^{\beta (E_i-\mu)} + 1 } \,,
  \end{equation}
  with the help of this theorem $(\beta = (k_{\rm B}T)^{-1})$.
  One gets a set of ten equations, which are linear in the unknown 
  correlation functions.  To solve this systems means that only
  properties of the cluster determine the correlations functions.

  However, in our special situation of $N=0$ each of the expectation values at
  the left hand side of (\ref{eq:spectraltheorem}) is supposed to vanish, 
  because they all include an 
  electron density operator. Then the solution of the set of equations becomes  
  trivial. One has to choose \( \mu \ll \varepsilon_k \forall k \), to fulfill the 
  requirements for the electron density. This leaves the set of spin correlation
  functions again undefined. 

  Hence, if $N=0$, the application of the spectral theorem to the 
  determined (that means electronic) set of Green's functions fails 
  to provide any information about spin correlation functions.
  Instead other Green's functions can be used. The equations of motion for 
  \( \GF{S_1^+}{S_1^-}, \GF{S_2^+}{S_2^-} \) and \( \GF{S_2^z S_1^+}{S_1^-} \) 
  are particularly simple in the case $N=0$. If the spectral theorem 
  is applied to its results, one obtains the information that 
  \( \CF{S_1^z}, \CF{S_2^z} \) and \( \CF{S_1^z S_2^z} \) are zero,
  respectively. However, even when using \( \GF{S_1^+}{S_2^-} \) or 
  \( \GF{S_2^+}{S_1^-} \) it is not possible to get any expression 
  for \( \CF{S_1^+ S_2^-} \) and \( \CF{S_1^- S_2^+} \).
  For these correlation functions the expectation value 
  has to be determined directly by calculating the trace of the
  canonical ensemble with the four possible eigenstates of $\bar {\cal H}$
  for $N=0$. The result is again zero. For an insulator it is not 
  surprising that all spin correlation functions vanish if they are
  solely determined within the cluster. 

 \subsection{Connection to the lattice}
 
  To create a connection of the two-site cluster to the lattice, the spin 
  correlation functions need to be specified outside the cluster. 
  It is the advantage of an insulator that this can be done without 
  coming into conflict with the spectral theorem
  applied to the cluster. 
 
  As explained in section \ref{Cluster model}
  we have quasi separated the electronic system and the spin system 
  by excluding \( {\cal H}_{f\!f} \) from the Hamiltonian. 
  As a different problem, we can now consider a situation of two 
  localized spins mutually interacting via \( {\cal H}_{f\!f} \) with
  \( J_{ij} = J_{\rm H} \) and embedded in the lattice via 
  an additional term in the Hamiltonian of the form
  \begin{equation}
    \label{eq:bfield}
    {\cal H}_f = - b \left( S_1^z + \eta S_2^z \right) .
  \end{equation}
  Simultaneous to the choice of $J_{\rm H}$, the sign parameter 
  $\eta$ allows to distinguish between ferromagnetic 
  $(\eta =+1)$ and antiferromagnet $(\eta=-1)$ configurations. In the 
  first case \( b = B \cdot \frac{g_J \mu_{\rm B}}{\hbar} \) is as a
  molecular or Weiss field, in the latter case it describes the crystal 
  anisotropy. The complete set of Green's functions for this problem 
  is derived in appendix \ref{The two-spin problem}. As a result 
  the expressions (\ref{ap:fm1})-(\ref{ap:fm4}) and 
  (\ref{ap:afm1})-(\ref{ap:afm4}) for a set of correlation functions 
  are achieved. 

  This set is identical with the correlation functions
  we need for the calculation of the electronic part described by
  \( \bar {\cal H} \). Determined in 
  this way and used in (\ref{eq:result}), we are sure that not only
  \( \CF{S^z_1} \) and \( \CF{S^z_2} \) but also the two-site
  correlation functions \( \CF{S^z_1 S^z_2}, \CF{S^+_1 S^-_2} \) and
  \( \CF{S^-_1 S^+_2} \) are described {\it consistently} by two experimentally
  accessible parameters $J_{\rm H}$ and $B$. We created an intelligible 
  connection to the lattice.

  Only after these considerations we are able to specify the behaviour
  of the electrons in a two-site (anti-)\-ferro\-mag\-netic Kondo lattice.
  If we assume a ferromagnet (\( \CF{S^z_1} = \CF{S^z_2} \)), we 
  obtain a dependence of the spectral weights belonging to the six energy 
  poles of the single-electron Green's function on the magnetization 
  as shown in figure \ref{fg:fm3d_1}.
  \begin{figure}[th]
    \begin{center}
     \epsfig{file=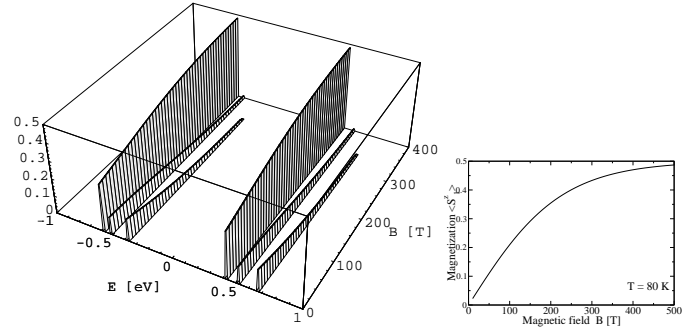, width=0.5\textwidth} 
    \end{center}
    \caption[Dependence of spectral weights on $B$]
    {For the energy poles of the single-electron Green's function, 
     the dependence of the corresponding spectral weights on a magnetization 
     \( \CF{S^z_1} = \CF{S^z_2} \) parallel to the spin of the test 
     electron $\ps$ is given (left). The latter is adjusted by an assumed 
     molecular field $B$ (right). The fixed model parameters are: 
     $J_{\rm H} \hbar^2 = 10^{-3} {\,\rm eV}$, $J \hbar = 0.3 {\,\rm eV}$,
     $\T = 0.5 {\,\rm eV}$ and $g_J = 1, T_0 = 0 {\,\rm eV}$. 
     The temperature is $80 {\,\rm K}$. 
    \label{fg:fm3d_1}}
  \end{figure}
  The magnetic field $B$ is uniquely used to create
  a ferromagnetic order of the system of localized spins and does not
  affect the conduction electrons directly. To obtain a saturation 
  of about 95\% one needs extremely large exchange fields of about
  $B = 400 {\rm\,T}$, as can also be estimated if the tendency
  to disorder, $k_{\rm B} T$, and to order, $\mu_{\rm B} B$, are set equal.
  One can clearly see that for these fields the energy poles
  $\Pmo$ and $\Ppo$ become dominant. This is not
  surprising, since the configuration where the excess electron and
  the two localized spins have all parallel spin has exactly this 
  excitation energy. The observation that for $B= 0{\,\rm T}$ these 
  two poles still possess the highest spectral weight is connected to
  the fact that additionally some one-electron eigenstates of the 
  Hamiltonian $\bar {\cal H}$ with the $z$-component of the total 
  spin $\pm (2S-1/2)$ have the same energy. The symmetry of lattice
  sites $1$ and $2$ nicely enters the figure, too: It is the reason 
  for similar structures in the groups around $E = -\T = -0.5 {\,\rm eV}$
  and around $E = \T = 0.5 {\,\rm eV}$.

  \begin{figure}[th]
    \begin{center}
     \epsfig{file=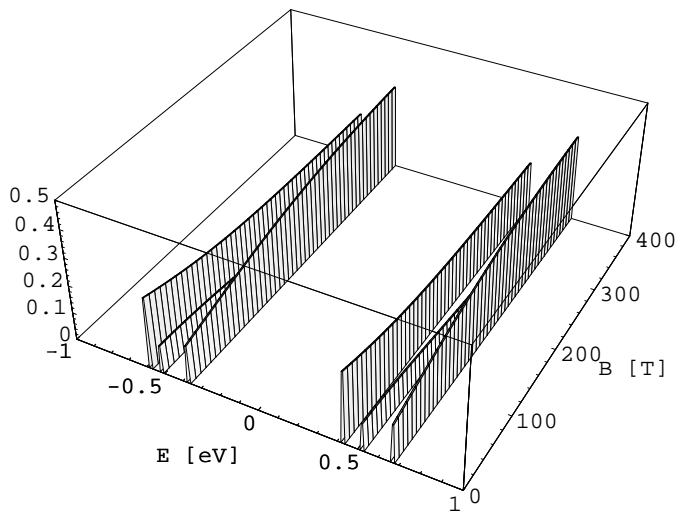, width=0.5\textwidth} 
    \end{center}
    \caption[Dependence of spectral weights on $B$]
    {Same as figure \ref{fg:fm3d_1} with the only difference
     that the spin of the test electron $\ps$ is antiparallel to the
     direction of the magnetization.
    \label{fg:fm3d_2}}
  \end{figure}
  As one can see in figure \ref{fg:fm3d_2} the situation is
  quite different if the spin of the test electron is antiparallel
  to the alignment of the localized spins, induced by the applied
  magnetic field. Here, the energy poles $\Ppwwp$ and $\Pmwwm$ get 
  maximum spectral weight when the field is increased. However,
  the site symmetry is not affected.

  If an antiferromagnetic configuration of the localized spins 
  (\( \CF{S^z_1} = -\CF{S^z_2} \)) is considered, there is again 
  no change in the positions of the energy poles (figure \ref{fg:poles}). 
  Of course this does not apply to their spectral weights. 
  \begin{figure}[th]
    \begin{center}
     \epsfig{file=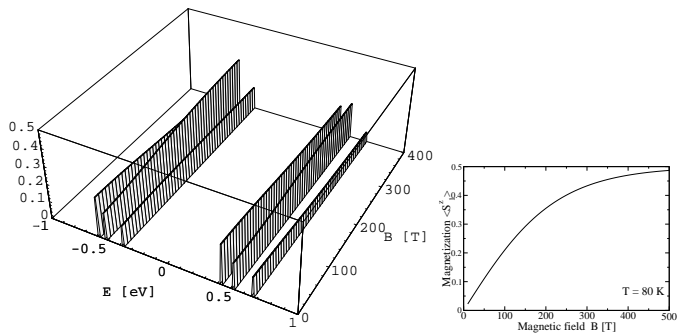, width=0.5\textwidth} 
    \end{center}
    \caption[Dependence of spectral weights on $B$]
    {Same as figure \ref{fg:fm3d_1}, but with 
     \( \CF{S^z_1} = -\CF{S^z_2} \) and $J_{\rm H} \hbar^2 =-10^{-3} {\rm\,eV}$. 
     $B$ is the magnetic anisotropy.
    \label{fg:afm3d_1}}
  \end{figure}
  Comparing figure \ref{fg:afm3d_1} to figure \ref{fg:fm3d_1}, one 
  notices already for zero magnetization ($\CF{S^z_1}=0 \Leftrightarrow B =0$)
  a different distribution. The opposite sign for $J_{\rm H}$ 
  affects the non-zero correlation functions $\CF{S^z_1 S^z_2}$ and
  $\CF{S^+_1 S^-_2}$. As concerns the dependence on $B$, primarily 
  the behaviour of the energy poles $\Ppwwm$ and $\Pmwwp$ has changed. 
  Their spectral weight does not tend to zero (ferromagnet), but 
  rather increases with $B$. In general the antiferromagnetic
  configuration leads to smaller changes of the spectral weights,
  since the average 
  \( \frac{1}{2} \left( \CF{S^z_1} + \CF{S^z_2} \right) \equiv 0 \)
  remains constant. 

  The term ${\cal H}_f$, which enters the determination of the 
  correlation functions, leads for $\eta=-1$ to a break-down of 
  the site symmetry. Therefore, the two groups of energy poles
  in figure \ref{fg:afm3d_1} show a slightly different dependence
  on $B$. One has to distinguish between 
  \( \GF{\cn{1\up}}{\cd{1\up}} \equiv \GF{\cn{2\dn}}{\cd{2\dn}} \) and
  \( \GF{\cn{1\dn}}{\cd{1\dn}} \equiv \GF{\cn{2\up}}{\cd{2\up}} \).
  The latter situation has been presented in figure \ref{fg:afm3d_2}. 
  \begin{figure}[th]
    \begin{center}
     \epsfig{file=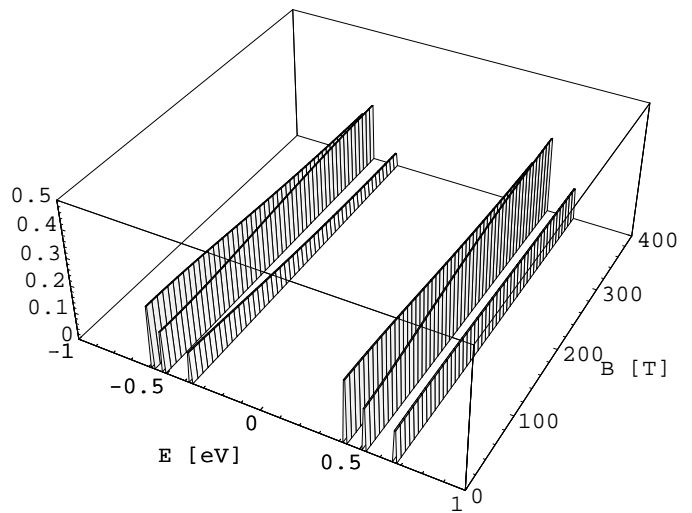, width=0.5\textwidth} 
    \end{center}
    \caption[Dependence of spectral weights on $B$]
    {Same as figure \ref{fg:afm3d_1}, but the Green's function
     is evaluated at the site which has a localized spin
     antiparallel to the electron spin.
    \label{fg:afm3d_2}}
  \end{figure}
  The dependence on the spin direction of the electron is of course
  much smaller than for a ferromagnetic configuration. 
  In fact, apart from an interchange of the two groups of energy poles the 
  distribution of the spectral weights shown in figure \ref{fg:afm3d_2} 
  is almost the same as in figure \ref{fg:afm3d_1}. 
  The formation of these two groups is a result of hybridization
  of the energy levels of the two single sites, as shown in
  figure \ref{fg:poles}. The fact that the spectral weights of the 
  energies in the lower group get the values of the higher group 
  (and vice versa) if the spin direction of the test electron is 
  changed, is again related to site symmetry.
  
  Of course, all these features can be explained by a detailed
  analysis of the eigenvalue problem. However, the Hilbert 
  spaces that need to be considered are relatively complex. 
  For instance the subspace corresponding to a $z$-component 
  of the total spin of $1/2$ is 6-dimensional, and hence every
  eigenvector is a linear combination of 6 different spin 
  configurations. Therefore, the possible analysis is too lengthy 
  to be presented here.

\section{Summary}

  We have shown that the Hamiltonian $\bar {\cal H}$ can be treated
  such that an analytic expression for the one-particle 
  Green's function as given in (\ref{eq:result}) is obtained. 
  In order to do that we formulated the complete set of equations
  of motion as a single matrix equation, made use of the symmetries 
  of the problem and combined Green's functions to reduce the matrix 
  to blocks of size $3 \times 3$, which can be solved. 
  In this article we studied the case of an insulator ($N=0$), but
  a generalization of these techniques to the situation in a metal
  is possible even so calculations are much less concise there.
  \cite{HRN01}

  Furthermore, the connection to a lattice has been discussed. 
  We propose to use the correlation functions in the spectral weights
  of the Green's functions for this purpose. A helpful tool is a
  two-site Heisenberg model which has been separated before from the
  Hamiltonian (\ref{eq:CKLM}) and calculated in appendix 
  \ref{The two-spin problem}. By adding a term ${\cal H}_f$ we
  can simulate ferromagnetic and antiferromagnetic lattices. 
  This assumption about the system of localized spins results in 
  certain features of the electronic partial system, visualized
  exemplary by the spectral weights given in figure \ref{fg:fm3d_1} 
  to \ref{fg:afm3d_2}. The site symmetry has a major influence
  on its distribution.

\section*{Acknowledgments}
 One of us (T.~H.) gratefully acknowledges the financial support of the 
 {\it Studienstiftung des deutschen Volkes}. This work also benefitted from the
 financial support of the {\it Sonderforschungsbereich SFB 290} of the 
 Deutsche Forschungsgemeinschaft.

\begin{widetext}
\begin{appendix}
 \section{The two-spin problem}
 \label{The two-spin problem}

 It is the purpose of this appendix to provide a consistent set of spin
 correlation functions as input for the two-site CKLM. This is done 
 by discussing the Heisenberg-type Hamiltonian
 \begin{equation}
   \label{ap:Hamiltonian}
   {\cal H}_{\rm A} 
    = - J_{\rm H} \left( S_1^+ S_2^- + S_1^- S_2^+ + 2 S_1^z S_2^z \right)
      - b \left( S_1^z + \eta S_2^z \right) . 
 \end{equation}
 To obtain correlation functions we use the spectral theorem,
 which needs certain Green's functions as an input. 
 Within an equation-of-motion (\ref{eq:eom}) approach these are:
 \begin{equation}
  \begin{array}{rclrcl}
    G_{11} &=& \GF{S_1^-}{S_1^+} & K_{11} &=& -2 \hbar^2 \CF{S_1^z} ,\\
    G_{21} &=& \GF{S_2^-}{S_1^+} & K_{21} &=& 0 ,\\
    \Gamma_{12} &=& \GF{S_1^z S_2^-}{S_1^+} \quad &
         L_{12} &=& \hbar^2 \CF{S_1^+ S_2^-} ,\\
    \Gamma_{21} &=& \GF{S_2^z S_1^-}{S_1^+} \quad &
         L_{21} &=& -2 \hbar^2 \CF{S_1^z S_2^z} .\\
  \end{array}
 \end{equation}
 On the right hand site we give the corresponding inhomogeneities 
 which appear in the equations of motion:
 \begin{eqnarray}
   \left( E + b \hbar \right) G_{11} 
   &=& K_{11} + 2 J_{\rm H} \hbar \left( \Gamma_{12} - \Gamma_{21} \right) \\
   \left( E + \eta b \hbar \right) G_{21} 
   &=& K_{21} - 2 J_{\rm H} \hbar \left( \Gamma_{12} - \Gamma_{21} \right) \\
   \left( E + \eta b \hbar \right) \Gamma_{12} 
   &=& L_{12} + J_{\rm H} \frac{\hbar^3}{2} \left( G_{11} - G_{21} \right) \\
   \left( E + b \hbar \right) \Gamma_{21} 
   &=& L_{21} - J_{\rm H} \frac{\hbar^3}{2} \left( G_{11} - G_{21} \right) 
 \end{eqnarray}
 Following the rules identical to what has been used in
 (\ref{eq:comb2}) the simplification
 \begin{eqnarray}
  \label{ap:comb} &
   \left(\begin{array}{cc}
     E + b \hbar \pm J_{\rm H} \hbar^2 & \mp J_{\rm H} \hbar^2 \\
     \mp J_{\rm H} \hbar^2 & E + \eta b \hbar \pm J_{\rm H} \hbar^2
   \end{array}\right)
   \left(\begin{array}{c}
     \fh G_{11} \pm \Gamma_{21} \\ \fh G_{21} \pm \Gamma_{12}
   \end{array}\right) 
  =\left(\begin{array}{c}
     \fh K_{11} \pm L_{21} \\ \fh K_{21} \pm L_{12}
   \end{array}\right)
 \end{eqnarray}
 is possible. The determinant of the emerging matrix, $D^{(\pm)}$, 
 has the form
 \begin{eqnarray}
   \label{ap:det}
   \frac{1}{ D^{(\pm)} } 
      &=& \frac{1}{2 \sqrt{ \ldots } }
          \left( \frac{1}{ E \pm J_{\rm H} \hbar^2 + \fh (1+\eta) b 
                           - \sqrt{ \ldots } }
               - \frac{1}{ E \pm J_{\rm H} \hbar^2 + \fh (1+\eta) b 
                           + \sqrt{ \ldots } }
          \right)  \nonumber \\ \mbox{with} \,\,
   \sqrt{ \ldots } 
      &=& \sqrt{ \left( J_{\rm H} \hbar^2 \right)^2 
                      + \frac{1}{2} (1-\eta) b^2 \hbar^2 } .
 \end{eqnarray}
 By taking the inverse of the matrix in (\ref{ap:comb}) and 
 after performing a partial fraction expansion one therefore 
 obtains the combined Green's functions 
 \begin{eqnarray*}
   \fh G_{11} \pm \Gamma_{21}
   &=& - \hbar^3 \CF{S_1^z} \left(
              \frac{(1-\eta)\frac{\hbar b}{4 \sqrt{\ldots}} + \frac{1}{2} }
                   { E - E_1^{(\pm)} } 
            - \frac{(1-\eta)\frac{\hbar b}{4 \sqrt{\ldots}} - \frac{1}{2} }
                   { E - E_2^{(\pm)} } 
       \right) \\
     &\mp& 2 \hbar^2 \CF{S_1^z S_2^z} \left(
              \frac{(1-\eta)\frac{\hbar b}{4 \sqrt{\ldots}} + \frac{1}{2} }
                   { E - E_1^{(\pm)} } 
            - \frac{(1-\eta)\frac{\hbar b}{4 \sqrt{\ldots}} - \frac{1}{2} }
                   { E - E_2^{(\pm)} } 
       \right) \\
     &-& \CF{S_1^+ S_2^-} \frac{J_{\rm H} \hbar^4}{2\sqrt{\ldots}}
       \left( \frac{1}{ E - E_1^{(\pm)} }
            - \frac{1}{ E - E_2^{(\pm)} }  \right) ,
               \\[2ex]
   \fh G_{21} \pm \Gamma_{12}
     &=& \pm \CF{S_1^z} \frac{J_{\rm H} \hbar^5}{2 \sqrt{ \ldots }} 
                \left( \frac{1}{ E - E_1^{(\pm)} } 
                     - \frac{1}{ E - E_2^{(\pm)} } \right) \\
       &+&  \CF{S_1^z S_2^z} \frac{J_{\rm H} \hbar^4}{\sqrt{ \ldots }} 
                \left( \frac{1}{ E - E_1^{(\pm)} } 
                     - \frac{1}{ E - E_2^{(\pm)} } \right) \\
       &\pm& \hbar^2 \CF{S_1^+ S_2^-} \left( 
              \frac{(\eta-1) \frac{\hbar b}{4 \sqrt{\ldots}} + \frac{1}{2} }
                   { E - E_1^{(\pm)} } 
            - \frac{(\eta-1) \frac{\hbar b}{4 \sqrt{\ldots}} - \frac{1}{2} }
                   { E - E_2^{(\pm)} }
              \right) 
 \end{eqnarray*}
 with the four energy poles
 \begin{eqnarray}
   E_1^{(\pm)} &=& \mp J_{\rm H} \hbar^2 - \fh (1+\eta) b 
         - \sqrt{ \left( J_{\rm H} \hbar^2 \right)^2 
                + \frac{1}{2} (1-\eta) b^2 \hbar^2 } , \nonumber \\
   E_2^{(\pm)} &=& \mp J_{\rm H} \hbar^2 - \fh (1+\eta) b 
         + \sqrt{ \left( J_{\rm H} \hbar^2 \right)^2 
                + \frac{1}{2} (1-\eta) b^2 \hbar^2 } . 
 \end{eqnarray}

 One can apply the spectral theorem (\ref{eq:spectraltheorem}) to
 these Green's functions. The corresponding left hand sides are
 \begin{equation}
  \begin{array}{lcrcl}
    G_{11} &\longrightarrow& 
    \CF{S_1^+ S_1^-} &=& \frac{\hbar^2}{2} +\hbar \CF{ S_1^z } ,\\
    G_{21} &\longrightarrow& 
    \CF{S_1^+ S_2^-} ,\\
    \Gamma_{12} &\longrightarrow& 
    \CF{S_1^+ S_1^z S_2^-} &=& - \fh \CF{S_1^+ S_2^-} ,\\
    \Gamma_{21} &\longrightarrow& 
    \CF{S_2^z S_1^+ S_1^-}
    &=& \frac{\hbar^2}{2} \CF{ S_2^z } +\hbar \CF{ S_1^z S_2^z } .
  \end{array}
 \end{equation}
 The emerging Bose distribution functions at the right hand side 
 of (\ref{eq:spectraltheorem}) can be abbreviated:
 \begin{equation}
   m_i^{(\pm)} = \frac{1}{{\rm e}^{\beta E_i^{(\pm)}} -1} .
 \end{equation}
 One ends up with a system of three equations which are linear
 in the correlation functions we are looking for. The fourth
 equation is just necessary to calculate \( \CF{S^z_2} \).
 It is most instructive to give the corresponding matrix 
 equations for each of the following two special situations
 separately.

 \subsection{Ferromagnet $(\eta=+1)$}

  Here, the expressions for the energy poles are
  \begin{equation}
  \begin{array}{rclrcl}
    E_1^{(+)} &=& - \hbar b - 2 J_{\rm H} \hbar^2 ,\quad&
    E_1^{(-)} &=& - \hbar b ,\\
    E_2^{(+)} &=& - \hbar b ,&
    E_2^{(-)} &=& - \hbar b + 2 J_{\rm H} \hbar^2 .
  \end{array}
  \end{equation}
  As a consequence \( m_1^{(-)} \equiv m_2^{(+)} \).
  The system of equations is:
  \[
  \left(\begin{array}{c}
      \frac{\hbar^2}{2} \\ 0 \\ 0
  \end{array}\right) =
  \left(\begin{array}{ccc}
    -1 - \frac{1}{2} m_3
    &
    - \frac{1}{2} \left( m_1^{(+)} - m_2^{(-)} \right)
    &  
    - \frac{1}{2} \left( m_1^{(+)} - m_2^{(-)} \right)
     \\
    \frac{1}{2} m_4
    &
    \frac{1}{2} \left( m_1^{(+)} - m_2^{(-)} \right)
    &
    -1 + \frac{1}{2} \left( m_1^{(+)} - m_2^{(-)} \right)
     \\
    \frac{1}{2} \left( m_1^{(+)} - m_2^{(-)} \right)
    & 
    \frac{1}{2} m_4
    & 
    1 + \frac{1}{2} m_3
  \end{array}\right)
  \left(\begin{array}{c}
    \hbar \CF{ S_1^z } \\
    2 \CF{S_1^z S_2^z} \\
    \CF{S_1^+ S_2^-}
  \end{array}\right) ,
 \]
  where the abbreviations 
  \( m_3 = m_1^{(+)} + 2 m_1^{(-)} + m_2^{(-)} \) and
  \( m_4 = m_1^{(+)} - 2 m_1^{(-)} + m_2^{(-)} \) have been used.

  This gives a solution for the correlation functions which can 
  be simplified to the expressions published before by Nolting \cite{NolM2}:
   \begin{eqnarray}
     \label{ap:fm1}
     \CF{S^z_1} 
     &=& \frac{\hbar}{2}
         \frac{\exp{\beta\hbar b}-\exp{-\beta\hbar b}}
              {1+\exp{-2\beta\hbar^2 J_{\rm H}}+\exp{\beta\hbar b}-\exp{-\beta\hbar b}},
      \\
     \label{ap:fm2}
     \CF{S^+_1 S^-_2} 
     &=& \frac{\hbar^2}{2}
         \frac{1-\exp{-2\beta\hbar^2 J_{\rm H}}}
              {1+\exp{-2\beta\hbar^2 J_{\rm H}}+\exp{\beta\hbar b}-\exp{-\beta\hbar b}},
      \\
     \label{ap:fm3}
     \CF{S^z_1 S^z_2}
     &=& \frac{\hbar^2}{4}
         \frac{\exp{\beta\hbar b}-\exp{-\beta\hbar b}-1-\exp{-2\beta\hbar^2 J_{\rm H}}}
              {1+\exp{-2\beta\hbar^2 J_{\rm H}}+\exp{\beta\hbar b}-\exp{-\beta\hbar b}},
      \\
     \label{ap:fm4}
     \CF{S^z_1} &=& \CF{S^z_2} 
   \end{eqnarray}
  In contrast to his calculations we obtain here the last line as a 
  result and not as an assumption. 

 \subsection{Antiferromagnet $(\eta=-1)$}

  Here, the expressions for the energy poles are
  \begin{eqnarray}
   E_1^{(+)} &=& - J_{\rm H} \hbar^2 - \sqrt{ \left( J_{\rm H} \hbar^2 \right)^2 +
                                      \left( b \hbar   \right)^2 } ,\\
   E_1^{(-)} &=& + J_{\rm H} \hbar^2 - \sqrt{ \left( J_{\rm H} \hbar^2 \right)^2 +
                                      \left( b \hbar   \right)^2 } ,\\
   E_2^{(+)} &=& - J_{\rm H} \hbar^2 + \sqrt{ \left( J_{\rm H} \hbar^2 \right)^2 +
                                      \left( b \hbar   \right)^2 } ,\\
   E_2^{(-)} &=& + J_{\rm H} \hbar^2 + \sqrt{ \left( J_{\rm H} \hbar^2 \right)^2 +
                                      \left( b \hbar   \right)^2 } 
  \end{eqnarray} 
  and consequently
  \begin{eqnarray*}
   \Longrightarrow \,
    m_1^{(+)} + m_2^{(-)} = -1 ,&\quad& 
    m_1^{(+)} - m_2^{(-)} = 2 m_1^{(+)} + 1 ,\\
    m_2^{(+)} + m_1^{(-)} = -1 ,&\quad&
    m_2^{(+)} - m_1^{(-)} = 2 m_2^{(+)} + 1 .
  \end{eqnarray*}
  The system of equations is:
 \[
  \left(\begin{array}{c}
      \frac{\hbar^2}{2} \\ 0 \\ 0
  \end{array}\right) =
  \left(\begin{array}{ccc}
       - \frac{\hbar b}{\sqrt{\ldots}}
         \left( m_1^{(+)} - m_2^{(+)} \right)
       &
       - \left( m_1^{(+)} + m_2^{(+)} + 1 \right)
       &
       - \frac{J_{\rm H} \hbar^2}{\sqrt{\ldots}}
         \left( m_1^{(+)} - m_2^{(+)} \right)
     \\
       0&
         \frac{J_{\rm H} \hbar^2}{\sqrt{ \ldots }} 
         \left( m_1^{(+)} - m_2^{(+)} \right)
       &
         \left( m_1^{(+)} + m_2^{(+)} \right)
     \\[2ex]
         \frac{J_{\rm H} \hbar^2}{\sqrt{ \ldots }} 
         \left( m_1^{(+)} - m_2^{(+)} \right)
       &0&
         \frac{\hbar b}{\sqrt{\ldots}} 
         \left( m_1^{(+)} - m_2^{(+)} \right)
     \\[2ex]
  \end{array}\right) \!\!
  \left(\begin{array}{c}
    \hbar \CF{ S_1^z } \\
    2 \CF{S_1^z S_2^z} \\
    \CF{S_1^+ S_2^-}
  \end{array}\right)  
 \]
  Here, the following set correlation functions, which to our
  knowledge has not been published before, is obtained.
 \begin{eqnarray}
   \label{ap:afm1}
   \CF{S_1^+ S_2^-} 
   &=& \frac{J_{\rm H}\hbar^2}{\sqrt{\left( J_{\rm H} \hbar^2 \right)^2 +
                             \left( b \hbar   \right)^2 }}
       \cdot \frac{\hbar^2}{2} \cdot
       \frac{ m_1^{(+)} - m_2^{(+)} }
            { m_1^{(+)} + 4 m_1^{(+)} m_2^{(+)} + m_2^{(+)} } \\[2ex]
   \label{ap:afm2}
   \CF{S_1^z S_2^z} 
   &=& - \frac{\hbar^2}{4} \cdot
       \frac{ m_1^{(+)} + m_2^{(+)} }
            { m_1^{(+)} + 4 m_1^{(+)} m_2^{(+)} + m_2^{(+)} } \\[2ex]
   \label{ap:afm3}
   \CF{ S_1^z } 
   &=& \frac{b \hbar}{\sqrt{\left( J_{\rm H} \hbar^2 \right)^2 +
                            \left( b \hbar   \right)^2 }}
       \cdot \frac{\hbar}{2} \cdot
       \frac{ m_1^{(+)} - m_2^{(+)} }
            { m_1^{(+)} + 4 m_1^{(+)} m_2^{(+)} + m_2^{(+)} } \\[2ex]
   \label{ap:afm4}
   \CF{ S_2^z } 
   &=& - \CF{ S_1^z } 
 \end{eqnarray}
  The last line, which characterizes an antiferromagnet, is again
  a result of the calculations.
\end{appendix}
\end{widetext}
\newpage$\mbox{ }$

\end{document}